\documentclass[sigconf]{acmart}

\usepackage{listings}
\lstdefinestyle{listingstyle}{
    basicstyle=\ttfamily\scriptsize,        
    breaklines=true,       
    frame=lines,
    numbers=none
}
\usepackage{enumitem}       
\usepackage{multirow}

\usepackage{soul}

\usepackage{xspace}
\newcommand{\likethis}{{$LikeThis!$}\xspace}

\usepackage{tabularx}
\newcommand{\centeringcell}[1]{\begin{tabular}{@{}c@{}}#1\end{tabular}}

\usepackage{xcolor}
\definecolor{revision}{rgb}{0,0,0}
\definecolor{strike}{rgb}{0.86,0.15,0.02}

\usepackage{xcolor}

\AtBeginDocument{%
  }

\copyrightyear{2026}
\acmYear{2026}
\acmConference[ICSE '26]{2026 IEEE/ACM 48th International Conference on Software Engineering}{April 12--18, 2026}{Rio de Janeiro, Brazil}
\acmBooktitle{2026 IEEE/ACM 48th International Conference on Software Engineering (ICSE '26), April 12--18, 2026, Rio de Janeiro, Brazil}
\acmPrice{}
\acmDOI{10.1145/3744916.3787834}
\acmISBN{979-8-4007-2025-3/2026/04}

\begin{document}

\title{$LikeThis!$ Empowering App Users to Submit UI Improvement Suggestions Instead of Complaints}

\author{Jialiang Wei}
\authornote{Both authors contributed equally to this research.}
\email{jialiang.wei@uni-hamburg.de}
\orcid{0009-0008-6028-1576}
\affiliation{
  \institution{University of Hamburg}
  \city{Hamburg}
  \country{Germany}
}

\author{Ali Ebrahimi Pourasad}
\authornotemark[1]
\email{ali.ebrahimi.pourasad@uni-hamburg.de}
\orcid{}
\affiliation{
  \institution{University of Hamburg}
  \city{Hamburg}
  \country{Germany}
}

\author{Walid Maalej}
\email{walid.maalej@uni-hamburg.de}
\orcid{0000-0002-6899-4393}
\affiliation{
  \institution{University of Hamburg}
  \city{Hamburg}
  \country{Germany}
}

\renewcommand{\shortauthors}{Wei, Ebrahimi Pourasad, and Maalej.}

\begin{abstract}
User feedback is crucial for the evolution of mobile apps.
However, research suggests that users tend to submit uninformative, vague, or destructive feedback.
Unlike recent AI4SE approaches that focus on generating code and other development artifacts, our work aims at empowering users to submit better and more constructive UI feedback with concrete suggestions on how to improve the app. 
We propose \likethis, a GenAI-based approach that takes a user comment with the corresponding screenshot to immediately generate multiple improvement alternatives, from which the user can easily choose their preferred option.
To evaluate \likethis, we first conducted a model benchmarking study based on a public dataset of carefully critiqued UI designs.
The results show that GPT-Image-1 significantly outperformed three other state-of-the-art image generation models in improving the designs to address UI issues while keeping the fidelity and without introducing new issues.
An intermediate step in \likethis to generate a solution specification before changing the design was key to achieving effective improvement.
Second, we conducted a user study with 10 production apps, where 15 users used \likethis to submit their feedback on encountered issues.
Later, the developers of the apps assessed the understandability and actionability of the feedback with and without generated improvements.
The results show that our approach helps generate better feedback from both user and developer perspectives, paving the way for AI-assisted user-developer collaboration.

\end{abstract}



\maketitle

\section{Introduction}

In the highly competitive app market, user satisfaction is a prerequisite for app success \cite{Martens:ReleaseEarlyRelease:2019,Lee:DeterminantsMobileApps:2014,Pagano:UserFeedbackAppstore:2013}.
User feedback thus represents a crucial direct channel for developers to understand user needs, identify and fix bugs, or gather suggestions for enhancements \cite{Hassan:EMSE:2018}.
App stores enable users to easily submit feedback as a text review and a 1-5 star rating.
However, this might sometimes be insufficient to convey the specific thoughts and ideas of users.
Research has shown that, while most reviews tend to be destructive or uninformative, many users are motivated to contribute in-depth, constructive feedback \cite{Pagano:UserFeedbackAppstore:2013,Haggag:LargeScaleAnalysis:2022,kurtanovic2017mining}.
Their ability to do so is, however, often constrained by language barriers in text communication, including mismatches in concepts and vocabulary between users and developers \cite{Zowghi:EASE:2015,haering2021automatically}. 
Even verbose user feedback can be too vague for developers to understand or might lack key details to act upon \cite{Maalej:AutomaticClassificationApp:2016,Martens:ExtractingAnalyzingContext:2019}.

Visual feedback pointing to the user interface (UI) helps communicate user needs and thoughts more precisely \cite{Maalej:WhenUsersBecomeCollaborators:2009}.
Visual mockups can convey specific ideas, e.g.~for a layout change or a new button with a clarity that is difficult to achieve with text alone.
While their benefits in requirements and design processes are well-documented, it remains challenging for users to easily create or edit UI mockups \cite{nielsen1994usability, Lee:GUIComp:2020}.
Powerful design tools, such as Figma \cite{figma} and Sketch \cite{sketch}, are built for professional designers and developers, not for average users, who simply want to easily and quickly share their feedback.

Beyond text and code generation \cite{Jiang:SurveyLargeLanguage:2025}, recent advance in Generative AI (GenAI) have made the creation of images \cite{Lei:IMAGINEEImageGeneration:2025,Huang:DiffusionModelBasedImage:2025} from text prompts very easy.
Researchers have already investigated the generation of UI from text prompts, but so far have focused only on generating new UIs instead of editing existing ones \cite{Wei:BoostingGUIPrototyping:2023,Wei:AIInspiredUserInterface:2025}.
Code-based UI generation \cite{Lu:MistyUIPrototyping:2025,Yuan:DesignRepairDualStreamDesign:2025} is restricted to specific UI technologies and requires the comprehension of code---thus targeting rather developers and professional designers. 
To the best of our knowledge, no work has investigated how to enable users to edit existing UIs while sharing in-situ feedback.

To address this gap, we propose \likethis, the first GenAI-based approach to empower users to directly submit improvement suggestions that address UI-related issues they encounter.
\likethis takes a brief user comment and a screenshot as input to first generate three alternative \textbf{design specifications} that address the issue encountered (``Suggestion Generation'').
Then, our approach edits the original UI image according to those specifications, generating three UIs that address the issue.
These images are then presented to the user, who can finally choose and submit the one that best fits their idea. 
We conducted two extensive studies to evaluate our approach: the first focuses on GenAI models' capabilities for UI improvement and the second on how effective and helpful is our approach for users and developers. 
Our research questions are as follows:

\begin{description}
    \item[RQ1:] How well can GenAI improve a UI that includes an issue?
        \item \textbf{RQ1.1:} How do state-of-the-art image generation \underline{models perform} in generating UI improvements addressing specific issues?
        \item \textbf{RQ1.2:} Does the use of \underline{masking} to point to a problematic area in the UI improve the results?
        \item \textbf{RQ1.3:} What is the impact of the intermediate \underline{``Suggestion}\newline\underline{Generation''} step of \likethis on the results?
    \item[RQ2:] Does our approach help create better user feedback?  
        \item \textbf{RQ2.1:} From the \underline{user perspective}, how accurate are the generated suggestions and how helpful is the whole approach?
        \item \textbf{RQ2.2:} From the \underline{developer perspective}, does \likethis improve the received feedback, particularly in terms of understandability and actionability?
\end{description}

To answer RQ1, we conducted a benchmarking study based on a public dataset of carefully evaluated mobile UIs \cite{Duan:UICritEnhancingAutomated:2024}.
We found that GPT-Image-1~\cite{gpt-image-1} clearly outperforms three other state-of-the-art image generation models in addressing UI issues, while maintaining high design fidelity and rarely introducing new issues.
Flux~\cite{flux-kontext}, Gemini~\cite{gemini_2_flash}, and Bagel~\cite{Deng:EmergingPropertiesUnified:2025} are able to address the issue, at least partly, only in 25-53\% of the evaluated screens.
When evaluating the impact of masking (i.e., highlighting the concerned region in the UI using a binary image with black and white), we observed that it only works well when the problematic area is small. 
For large areas, masking often negatively impacts fidelity and robustness to new issues.
We also found that our intermediate ``Suggestion Generation'' step was crucial for achieving top performance.
To answer RQ2, we asked 15 users to use 10 production apps, identify issues, and submit feedback using \likethis. 
Participants agreed that 85.5\% of generated suggestions were mostly or very accurate.
The corresponding app developers assessed the submitted feedback, confirming that our approach improves the understandability and actionability of the feedback, helping to gather higher-quality user feedback.

We introduce our approach and an iOS implementation in Section~\ref{sec:appraoch}.
Then, Section~\ref{sec:model-study} reports on the model benchmarking study followed by the users' and developers' study in Section~\ref{sec:user-study}.
We discuss the threats to validity and limitations in Section~\ref{sec:threats} and related work in Section~\ref{sec:rel-work}.
Finally, Section~\ref{sec:conlusion} sketches the road ahead for GenAI-assisted user-developers interaction.
We share the data and scripts of both studies in our replication package \cite{replicationPackage}.

\section{Generating User Feedback with \likethis}
\label{sec:appraoch}
We propose \likethis, a novel approach designed to empower users to submit more constructive feedback by leveraging multimodal large language models (MLLMs) and image generation models.
The approach takes textual user feedback and an app's screenshot as input and generates multiple UI design suggestions that directly address the reported issue.
The user can then select their preferred suggestion to submit together with their feedback.
For instance, instead of submitting a vague complaint like ``This text is hard to read'', a user can select from AI-generated design suggestions that, for example, ``increase the font size'' or ``adjust the contrast''.
This makes it clear to developers what exactly the user finds problematic, turning vague feedback into a concrete and actionable suggestion.
We present the workflow of \likethis and a prototype for iOS.

\subsection{Workflow} \label{sec:workflow}

\begin{figure}[]
\centering
\includegraphics[width=\linewidth]{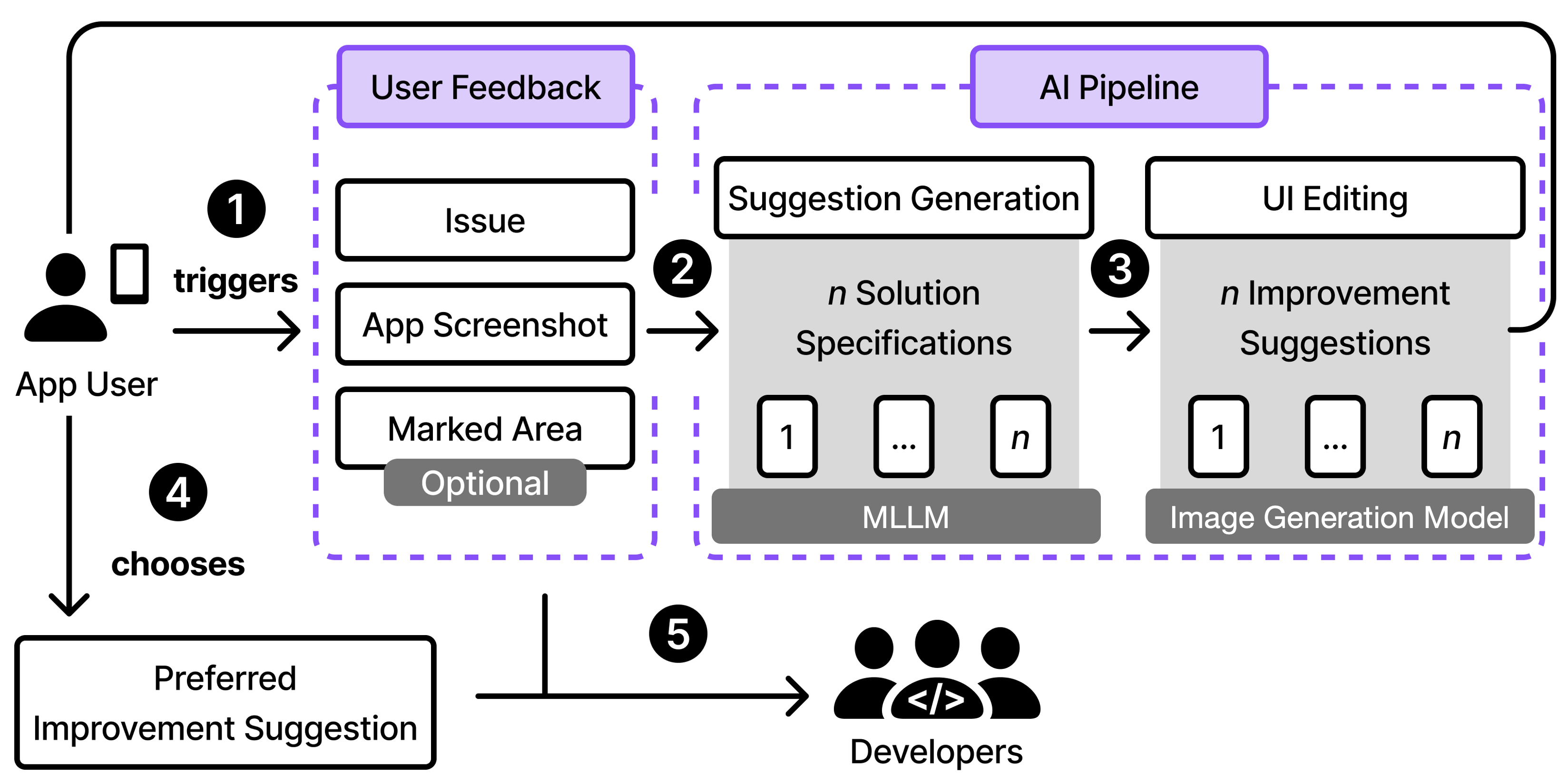}
\caption{
Overview of \likethis.
A user reports an app issue along with a screenshot (1).
The AI pipeline generates \textit{n} solution specifications (2), which are used to create improvement suggestions (3).
The user selects their preferred suggestion (4) and submits it along with the issue to developers (5).
}
\label{fig:workflow}
\end{figure}

Figure~\ref{fig:workflow} shows an overview of \likethis.
Users can report an issue by submitting a text comment and a screenshot of the app screen where the issue occurs (1).
They can also mark areas indicating the problematic UI part. 
For example, they may highlight text they find hard to read.
This marked area is converted into a mask image, which is commonly used by image generation models to guide the editing process~\cite{Rombach:HighResolutionImageSynthesis:2022}, ensuring changes are localised in the marked area while preserving the surrounding UI elements.
This feedback is then processed through an AI pipeline including 2 steps.
The first step, ``Suggestion Generation'', creates \textit{n} different textual solution specifications that describe how the UI could be changed to address the user issue (2).
The second step, ``UI Editing'', applies these \textit{n} instructions to the original screenshot to generate \textit{n} UI improvement suggestions (3).
In other words, these images show different alternatives of how the UI could look once the issue is resolved.
After generating multiple UI improvement suggestions, the user chooses their preferred option (4), which is then submitted to the development team together with the initial feedback (5).

\subsubsection{Suggestion Generation}
Recent studies have highlighted the capabilities of MLLMs \cite{Wu:MultimodalLargeLanguage:2023,Caffagni:RevolutionMultimodalLarge:2024} in understanding mobile UIs, such as detecting UI issues \cite{Pourasad:DoesGenAIMake:2025,Duan:UICritEnhancingAutomated:2024,Xiang:SimUserGeneratingUsability:2024}, automated UI testing \cite{Liu:MakeLLMTesting:2024,Feng:AgentUserTesting:2025}, and automating app tasks \cite{Hong:CogAgentVisualLanguage:2024,Wang:MobileAgentv2MobileDevice:2024,Zhang:AppAgentMultimodalAgents:2025}. 
We leverage MLLMs for our ``Suggestion Generation'' step. 
We employ Prompt \ref{prompt:solution}, which takes user feedback and an app screenshot as input.
The MLLM (e.g., GPT-4o), then generates a list of $n$ solution specifications, each containing a title and a detailed description.
Additionally, a UI description of the screenshot is generated for the subsequent step.

\subsubsection{UI Editing}
The ``UI Editing'' is based on image generation models \cite{Huang:DiffusionModelBasedImage:2025} (e.g., GPT-image-1), a family of models that can edit a given image based on a given instruction.
For this step, we instruct the model with Prompt \ref{prompt:ui-edit} using the user feedback, the UI description and solution specification obtained in the previous step.
This prompt, along with the original screenshot and an optional mask, serves as input to the image generation model to generate the final suggestions of improved UIs.

We initially employed Prompt \ref{prompt:ui-edit-no-solution} to directly generate UI modifications from user-reported issues.
However, pilot testing revealed that this approach often produced incomplete or irrelevant edits. 
To address this problem, we introduced the intermediate ``Suggestion Generation'' step.
In an ablation study (Section~\ref{sec:prompt-eval}), we compare this dedicated two-step workflow with the initial direct-generation approach.
The final version of our prompts also includes refinements derived from multiple pilot iterations, such as instructing the LLM to avoid describing animations and to focus strictly on static UI changes.

\begin{lstlisting}[float, numbers=none, caption={Prompt for Suggestion Generation}, label={prompt:solution}]
# Task
You are a UI/UX design expert. Your objective is to analyze the given UI image and propose {n} design modifications that address the user feedback provided below.

For each modification, provide a highly detailed description of the proposed UI in a single, static state. Focus on: layout, UI components, labels, and visual hierarchy.

# Important constraints
- Do not include any visual mockups or text-based wireframes.
- Do not describe animations, transitions, or other dynamic behaviors.

# User feedback:
{user_feedback}

# Output format:
```json
{
"ui_description": "Concise summary of the current UI and its purpose.",
"modifications": [
    {
      "title": "A concise title of the proposed modification.",
      "description": "A detailed description of the proposed modification."
    }
  ]
}
```
\end{lstlisting}

\begin{lstlisting}[float, numbers=none, caption={Prompt for UI Editing}, label={prompt:ui-edit}]
# Context
You are provided with a screenshot of a mobile app along with the following description:
{ui_description}

# User Feedback
Users reported the following issue:
{user_feedback}

# Task
As a UI/UX expert, modify the screen according to the following instruction so that the final design fully resolves the user feedback, while preserving all other visual elements of the original design:

{modification_title}, which can be achieved by: {modification_description}
\end{lstlisting}

\begin{lstlisting}[float, numbers=none, caption={Prompt for UI Editing without ``Suggestion Generation''}, label={prompt:ui-edit-no-solution}]
"""
# User Feedback
Users reported the following issue:
{user_feedback}

# Task
As a UI/UX expert, modify the screen to fully resolves the above user feedback, while preserving all other visual elements of the original design.
"""
\end{lstlisting}

\subsection{iOS Implementation}

\begin{figure*}
\centering
\includegraphics[width=1.0\textwidth]{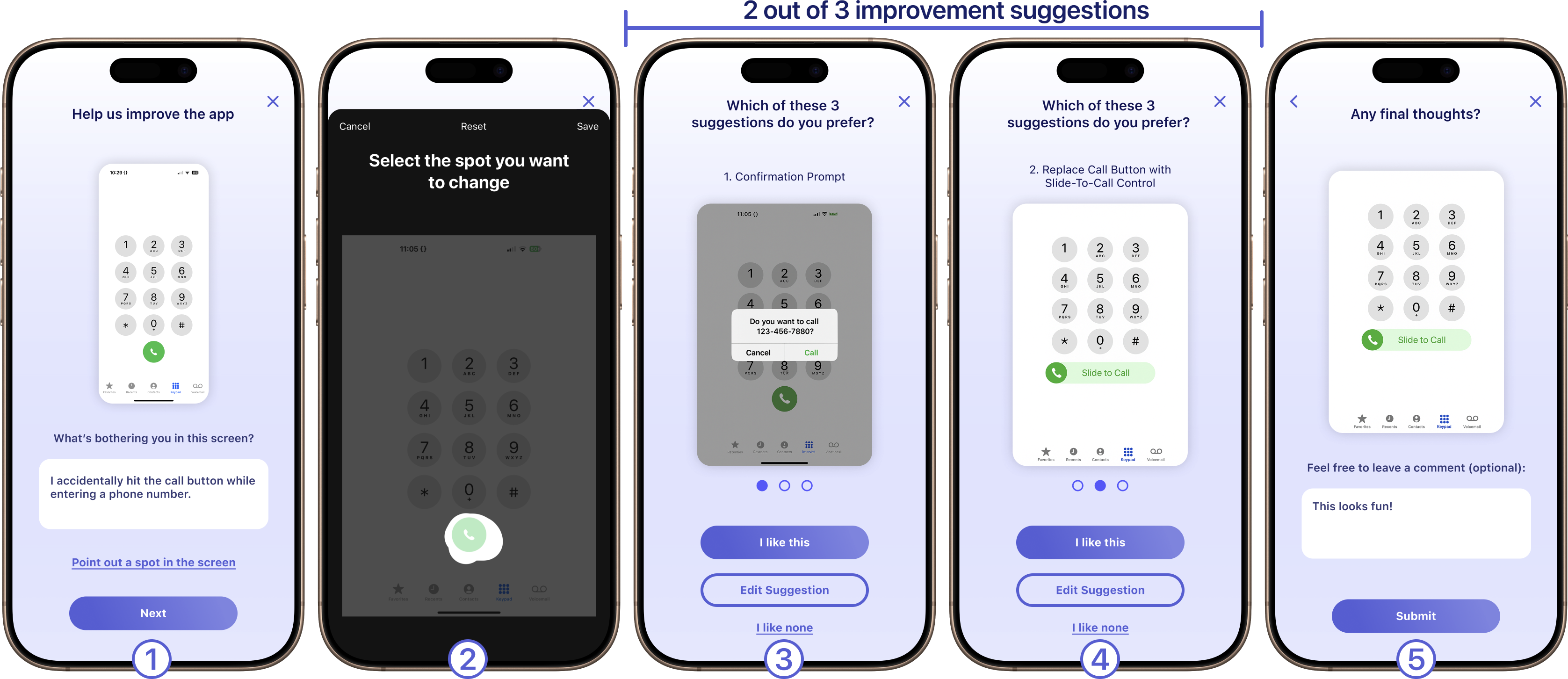}
\caption{\likethis iOS app enables users to report issues by capturing a screenshot, describing their issue, and optionally marking an area (Screens 1–2). Here, the user reports accidentally tapping the call button while entering a number. \likethis generates improvement suggestions (Screens 3–4). Users can then select a suggestion to submit, edit, or reject all. In this case, the user chooses to change the call button to a Slide-To-Call control and submits it with an optional comment (Screen 5).}
\label{fig:ios-approach}
\end{figure*}

We implemented an iOS prototype of \likethis as shown in Figure~\ref{fig:ios-approach}.
When users encounter an issue, e.g., unreadable text, they can take a screenshot and upload it.
They can then describe the issue in a text field, as shown in Screen~1.
In this example, a screenshot from the phone app was uploaded with the issue: ``I accidentally hit the call button while entering a phone number''.
Users can optionally highlight a specific area, as shown in Screen~2, where the call button is marked.
After submitting the issue, the app displays a loading screen.
When generation is complete, users are presented with three improvement suggestions.
In our example, two are shown in Screen~3 and~4.
Screen~3 presents a solution where a confirmation dialogue appears before initiating a call.
In screen~4, the proposed solution replaces the call button with a Slide-To-Call control.
At this point, the user can either accept one of the suggestions, edit a specific suggestion, or reject all of them.
When editing, users can describe desired changes in a screen similar to Screen~1, which generates a new suggestion that can be iteratively refined.
After choosing a suggestion, the users can submit it along with an optional comment, as shown in Screen~5.
The final report includes the original app screenshot, a marked screenshot (optional), the user-reported issue, the selected improvement suggestion, and a final comment (optional).

Our iOS implementation is built in \texttt{SwiftUI} and available as a Swift Package, making it easy for app developers to integrate into their projects~\cite{apple_swift_packages_doc}.
For the ``Suggestion Generation'', the app uses GPT-4o with a temperature of 1.
For the ``UI Editing'', it uses GPT-Image-1, which is selected due to its superior performance shown in Section \ref{sec:model-eval}.
Our implementation includes an MLLM service layer, enabling seamless model replacement by the app developers.

\section{Models Benchmarking Study}
\label{sec:model-study}
State-of-the-art image generation models have demonstrated strong capabilities in general image editing, but how well do they perform when editing mobile UI images?
To address RQ1, we evaluated the performance of various image generation models on a set of 300 UI images randomly selected from the UICrit dataset \cite{Duan:UICritEnhancingAutomated:2024}.
Two annotators independently assessed the edited UI images based on four criteria: \textit{User Preference Ranking}, \textit{Issue Resolution}, \textit{Fidelity to Original}, and \textit{Robustness to New Problems}, as explained in the following.

\subsection{Dataset Preparation} \label{sec:dataset}
The evaluation requires a dataset with mobile UI images, user feedback on the UI, and a mask highlighting the area to edit (only for models that supports this feature).
The UICrit dataset satisfy these requirements \cite{Duan:UICritEnhancingAutomated:2024}. 
It comprises 3,059 design critiques for 983 mobile UI screenshots, collected from seven professional designers.
The UICrit dataset was constructed by randomly sampling 1,000 screenshots from the CLAY dataset \cite{Li:LearningDenoiseRaw:2022}, a cleaned subset of the RICO dataset \cite{Deka:RicoMobileApp:2017}.
Seven designers were then tasked with identifying issues in the sampled screenshots.
For each identified issue, they drew a bounding box around the problematic area and provided a structured textual critique. 
Each critique includes (1) the expected design standard from related guidelines \cite{Nielsen:EnhancingExplanatoryPower:1994,Luther:StructuringAggregatingEvaluating:2015,apple-design-guideline}, (2) the issue, and (3) suggestions for resolving the issue. 
For instance, a critique might state: ``The expected standard is that text should be easy to read. 
In the current design, the text is too small.
To fix this, increase the font size.''

\begin{figure}[!htp]
\centering
\includegraphics[width=1\linewidth]{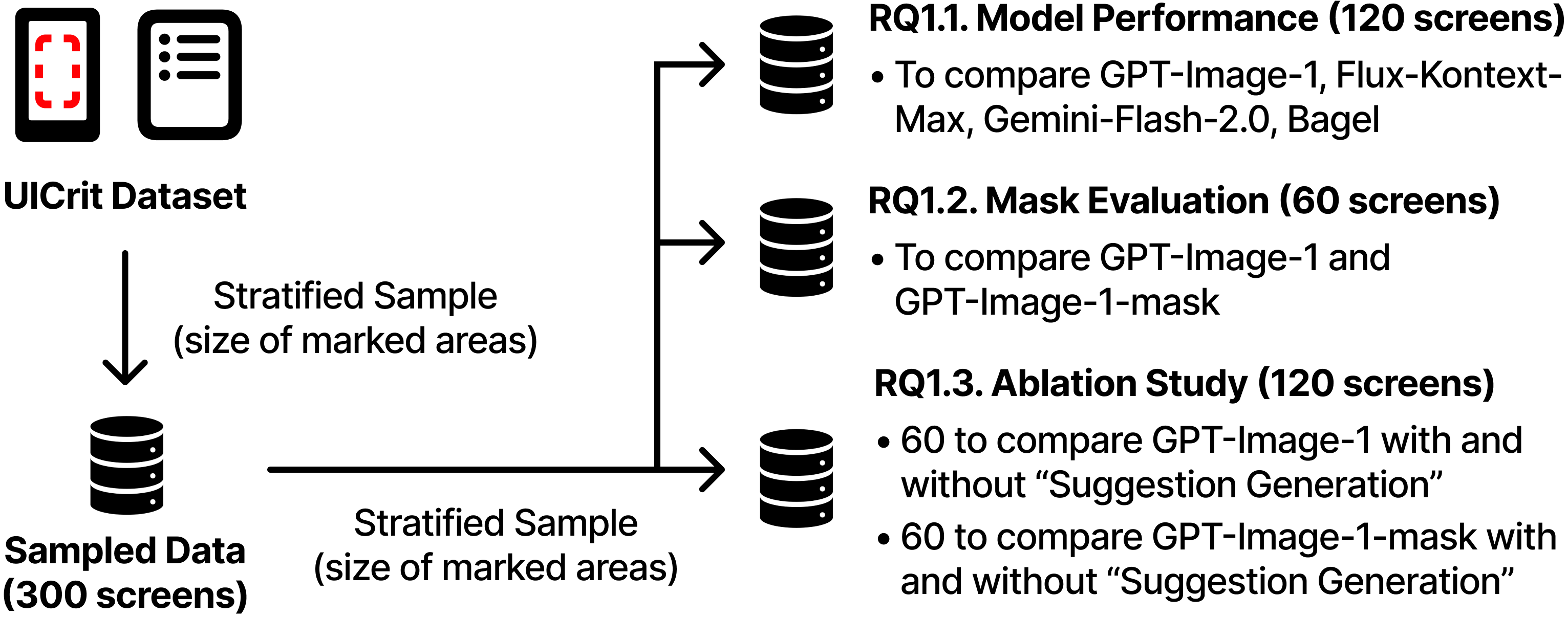}
\caption{Data sampling from UICrit dataset for RQ1.}
\label{fig:dataset}
\end{figure}

To answer RQ1, we randomly sampled 300 critiques/screens from the UICrit dataset as illustrated in Figure \ref{fig:dataset}.
We constructed a stratified sample, ensuring each critique is from a distinct screenshot, as one screenshot may have multiple critiques.
Stratification was based on the proportion of bounding box area relative to the entire screenshot, divided into three intervals: 0\%-20\%, 20\%-80\%, 80\%-100\%.
The goal of this stratification is to ensure the inclusion of various types of issues covering small, medium, or full-screen areas.
In the UICrit dataset, most of the bounding box areas represent a small proportion of the screen (<20\%), while few fall within 20\%-80\%. 
Using narrower split would result in some ranges with too few samples for stable comparison.
We selected the second sentence of each critique (which always describes the issue) as user feedback.
The first sentence (which states the design guideline) and the third sentence (which suggests a modification) were excluded, as typical user feedback is unlikely to include such information \cite{Pagano:UserFeedbackAppstore:2013}.
It is worth noting that the third sentences were not used as ground truth for the ``Suggestion Generation'' either, since issues can have multiple valid fixes \cite{Terry:VariationElementAction:2004}.
Our resulting dataset contains 300 unique screenshots, each paired with user feedback and an associated mask area (bounding box).
This provides the necessary input for the prompts of ``Suggestion Generation'' (Prompt~\ref{prompt:solution}) and ``UI Editing'' (Prompt~\ref{prompt:ui-edit}).
Instead of reusing the same 300 screenshots across RQ1.1-1.3, we split them in the same stratified manner (Figure \ref{fig:dataset}), yielding three distinct, non-overlapping subsets for each research question.
This design ensures that no interface is ``seen'' in more than one experiment, eliminating carry-over effects  \cite{Hornbaek:WhysHowsExperiments:2013}.
At the end, we created three splits: 120 screenshots to evaluate model performance (RQ1.1, Section~\ref{sec:model-eval}), 60 screenshots to assess the impact of masking (RQ1.2, Section~\ref{sec:mask-eval}), and another 120 screenshots for the ablation study (RQ1.3, Section~\ref{sec:prompt-eval}).

\subsection{Evaluation Protocol} \label{sec:eval-protocol}

\begin{figure}[!htp]
\centering
\includegraphics[width=\linewidth]{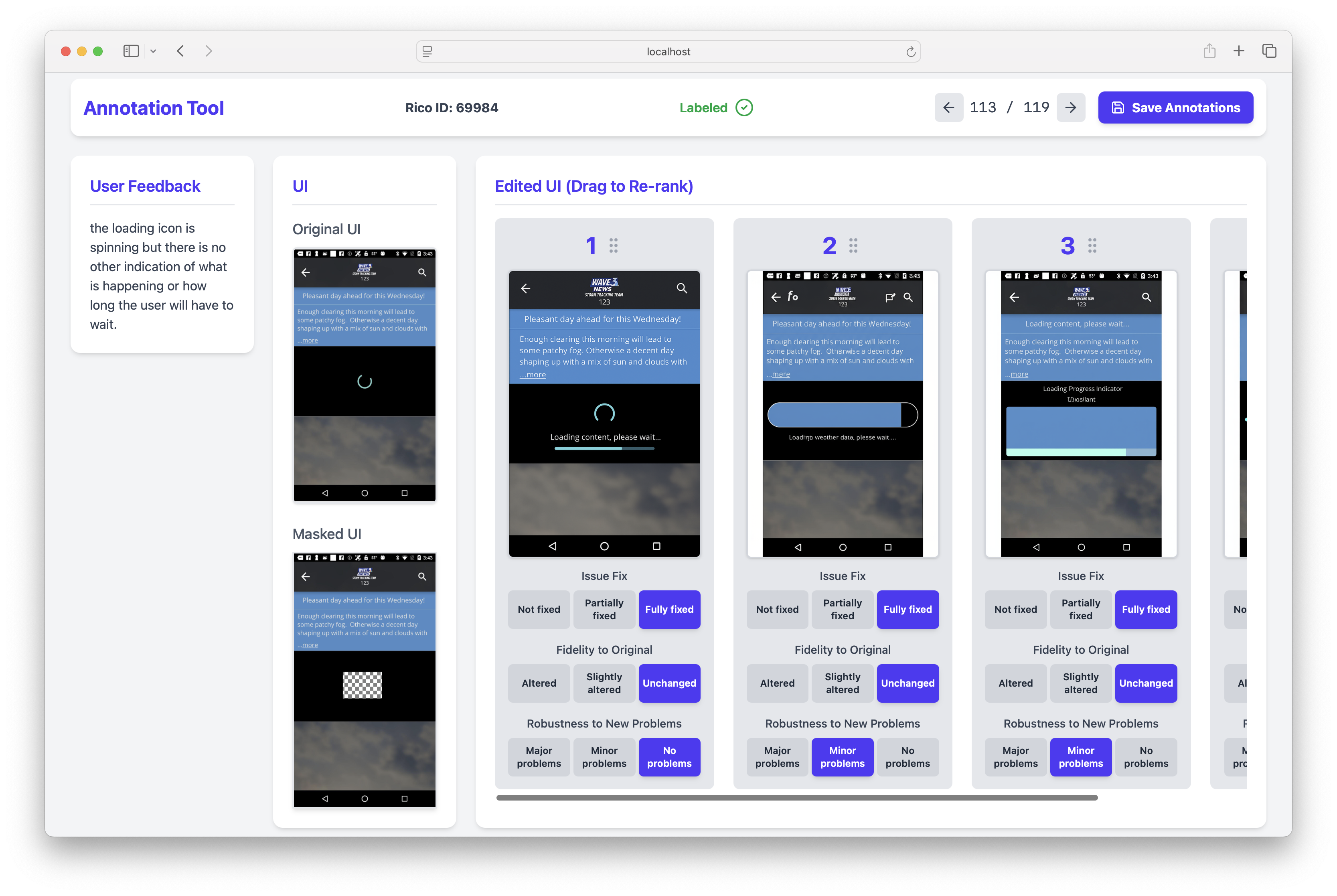}
\caption{Evaluation tool to assess generated UIs.}
\label{fig:annotation-tool}
\end{figure}

To facilitate the comparison of different models, we developed an annotation tool illustrated in Figure~\ref{fig:annotation-tool}. 
The first column of the tool displays the user feedback used as input for the ``Suggestion Generation'' step.
The second column presents the visual prompt, it contains both the original UI and the corresponding masked UI. 
The remaining columns show the UIs generated by different models or under different settings.
To mitigate potential evaluator bias, the tool randomly shuffles the order of these generated UIs, making it difficult to infer the model that created each UI. 
Additionally, all UIs were padded with white to a 2:3 aspect ratio, ensuring that image dimensions do not reveal the generating model.
Building on evaluation metrics commonly used in image editing research (e.g., editing accuracy and contextual preservation \cite{Huang:DiffusionModelBasedImage:2025}), we defined the following four metrics to assess the generated UIs.

\textbf{User Preference Rank:}
The relative ranking of the generated UIs, as evaluated by human.
The evaluators were instructed to assume the role of a user reporting the issues, and rank the UIs according to their preference. 
A lower numerical ranking indicates better UI.


\textbf{Issue Resolution:}
This metric assesses how well does the UI resolves user-reported issues. 
Scores are: 1 for issue "Not fixed", 2 for "Partially fixed", and 3 for "Fully fixed".
A higher score indicates better issue resolution.


\textbf{Fidelity to Original:}
This metric assesses how well the UI elements not targeted by the feedback remained unchanged.
Scores are: 1 for "Altered", 2 for "Slightly altered", and 3 for "Unchanged".
A higher score indicates greater fidelity.


\textbf{Robustness to New Problems:}
This metric captures any new problems introduced in the generated UI, such as layout breakage, novel bugs, or worsened usability.
Scores are: 1 for "Major problems", 2 for "Minor problems", and 3 for "No problems".
A higher score indicates fewer introduced problems.

For evaluation, we followed a common annotation protocol \cite{Krippendorff:ContentAnalysisIntroduction:2018} including four steps.
First, three authors, each with at least five years of software development experience, jointly developed a detailed annotation guideline, including precise definitions of the four evaluation metrics. 
Second, the first two authors independently annotated a test sample of 15 tasks (not included in the aforementioned 300) using the annotation tool.
Third, they met to discuss and resolve any disagreements and ensure a shared understanding of the guideline.
Finally, both authors independently annotated the evaluation samples.
Their scores were averaged to obtain the final results.

\subsection{RQ1.1. Model Performance} \label{sec:model-eval}

We benchmarked four state-of-the-art models released in 2025 to investigate their performance in the context of UI editing.

\subsubsection{Evaluated Models}

We compared GPT-Image-1 \cite{gpt-image-1}, Flux-Kontext-Max \cite{flux-kontext}, Gemini-2.0-Flash \cite{gemini_2_flash}, and Bagel \cite{Deng:EmergingPropertiesUnified:2025}, which were selected due to their strong performance in image editing tasks \cite{image_leadboard}. 
Each model was evaluated using both the UI image and a textual prompt as inputs. 

\vspace{0.2em}

\textbf{GPT-Image-1} \cite{gpt-image-1} is OpenAI’s latest multimodal image generation model, capable of producing high-quality images from both textual and visual inputs. 
For our experiments, we configured the model with the ``quality'' parameter set to ``high''.

\vspace{0.2em}

\textbf{Flux-Kontext-Max} \cite{flux-kontext} is from a family of generative models designed to unify image generation and editing tasks. 
These models offer competitive performance with improved inference speed relative to other state-of-the-art models. 
We used Flux-Kontext-Max, which is the highest-performing variant, with the ``aspect\_ratio'' parameter set to ``match\_input\_image''.

\vspace{0.2em}

\textbf{Gemini-2.0-Flash} \cite{gemini_2_flash} is a member of the Gemini 2.0 series, a suite of highly-capable models that build upon the sparse Mixture-of-Experts (MoE) Transformer architecture.
We used the gemini-2.0-flash-preview-image-generation API from Google's Vertex AI for our work.

\vspace{0.2em}

\textbf{Bagel} \cite{Deng:EmergingPropertiesUnified:2025} is an open-source, decoder-only foundation model developed by ByteDance that supports multimodal understanding and generation. 
We employed Bagel in its ``image-editing'' mode, enabling the ``enable\_thinking'' parameter and setting ``output\_quality'' to the max value.

\subsubsection{Evaluation Method}

We used the screenshots and their corresponding feedback collected in Section \ref{sec:dataset} as input for Prompt \ref{prompt:solution} to generate \textit{one} solution specification using GPT-4o, and used its output to construct Prompt \ref{prompt:ui-edit}. 
The latter prompt, along with the original screenshot, was subsequently used by four evaluated model.
This process yielded a total of 480 generated UI images based on the 120 original screenshots, referred to as 120 tasks.
Finally, two annotators independently annotated these 120 tasks with our annotation tool, following the guideline established in Section~\ref{sec:eval-protocol}.

\begin{table}[]
\caption{Results of 240 annotations (by 2 annotators for 120 tasks) comparing four models.}

\scalebox{0.88}{
\tabcolsep=0.1cm
\begin{tabular}{l | c | ccc|c | ccc|c | ccc|c}
\hline
 & \textbf{F\#1} & \multicolumn{4}{c|}{\textbf{Resolution}} & \multicolumn{4}{c|}{\textbf{Fidelity}} & \multicolumn{4}{c}{\textbf{Robustness}} \\
 &  & \multicolumn{3}{c|}{\textbf{Frequency}} & \textbf{Avg} & \multicolumn{3}{c|}{\textbf{Frequency}} & \textbf{Avg} & \multicolumn{3}{c|}{\textbf{Frequency}} & \textbf{Avg} \\
 &  & \textbf{1} & \textbf{2} & \textbf{3} &  & \textbf{1} & \textbf{2} & \textbf{3} &  & \textbf{1} & \textbf{2} & \textbf{3} &  \\
\hline
GPT & \textbf{214} & 22 & 18 & 200 & \textbf{2.74} & 4 & 33 & 203 & 2.83 & 1 & 27 & 212 & \textbf{2.88} \\
Flux & 14 & 153 & 29 & 58 & 1.60 & 4 & 30 & 206 & 2.84 & 11 & 70 & 159 & 2.62 \\
Gemini & 8 & 114 & 53 & 73 & 1.83 & 18 & 43 & 179 & 2.67 & 35 & 115 & 90 & 2.23 \\
Bagel & 4 & 181 & 27 & 32 & 1.38 & 9 & 19 & 212 & \textbf{2.85} & 100 & 35 & 105 & 2.02 \\
\hline
\end{tabular}
}

\vspace{0.5em}

\raggedright
\footnotesize
\textbf{F\#1}: Frequency of achieving rank 1 in \textit{User Preference Rank}; 
\textbf{Frequency} of the scores 1,2,3;
\textbf{Avg}: Average score;
\textbf{Resolution}: \textit{Issue Resolution}; 
\textbf{Fidelity}: \textit{Fidelity to Original}; 
\textbf{Robustness}: \textit{Robustness to New Problems}.

\label{tab:model-eval}
\end{table}

\subsubsection{Evaluation Result}

Table \ref{tab:model-eval} shows the performance of the four evaluated models, GPT-Image-1, Flux-knotext-max, Gemini-2.0-Flash, and Bagel. 
The evaluation involved 120 tasks, with data aggregated from two annotators (80.42\% agreement).

GPT-Image-1 performed best in \textit{User Preference Rank} (214 vs. 4-14 for other models), \textit{Issue Resolution} (2.75 vs. 1.37-1.83 for others), and \textit{Robustness to New Problems} (2.89 vs. 2.03-2.62 for others), with all results being statistically significant (Wilcoxon-Mann-Whitney test $p\leq0.05$).
Flux, Gemini, and Bagel showed some potential for \textit{Issue Resolution}, but only on 25-53\% of the screens samples.
While Flux demonstrated good Robustness, and all models achieve comparable performance in \textit{Fidelity}, GPT-Image-1's overall performance was superior.

\subsection{RQ1.2. Impact of Masking} \label{sec:mask-eval}
The previous evaluation highlights GPT-Image-1's superior performance in generating improved UIs. 
Unlike the other three evaluated models, GPT-Image-1 also supports image inpainting.
This feature allows GPT-Image-1 to accept an optional mask image in addition to a textual prompt and the original image.
The mask highlights editable areas, enabling the model to focus its edits on specific regions and facilitating controlled image modification.
Here we will investigate the impact of the mask in UI improvement.

\subsubsection{Evaluation Settings}
We evaluated GPT-Image-1 under two distinct input conditions: with and without an input mask, where the former referred to as GPT-Image-1-mask. 
All parameters, except for the mask input, remained identical in both settings.

\subsubsection{Evaluation Method}
During evaluation, we used the 60 screenshots and corresponding user feedback collected in Section \ref{sec:dataset}.
The UI generation process for GPT-Image-1 remained consistent with Section \ref{sec:model-eval}. 
For GPT-Image-1-mask, we used the bounding boxes from the UICrit dataset directly as mask inputs.
This procedure resulted in a total of 120 generated UI images derived from the 60 original screenshots, referred to as 60 tasks.
Then, two annotators annotated the 60 tasks independently with our annotation tool following the same guideline from Section \ref{sec:eval-protocol}.
Again, the annotators did not know which image is generated by which model.

\subsubsection{Evaluation Result}

As Table \ref{tab:mask-eval} shows, the performance of using a mask with GPT-Image-1 is highly dependent on the proportion of the bounding box area relative to the entire screenshot. 
The two annotators had an agreement rate of 82.5\% across the 60 tasks.

For small-area modifications (0\%-20\% of the entire screen), using a mask is highly beneficial. 
The masked condition achieved a superior \textit{User Preference Rank} (28 vs. 12) and perfect fidelity (3.00 vs. 2.90), while also slightly better in \textit{Issue Resolution}, all with $p\leq0.05$.
The unmasked condition was only marginally better at not introducing new problems. 

Conversely, for medium (20\%-80\%) to large (80\%-100\%) -affected area, the benefit of masking diminishes, and the unmasked approach generally becomes more effective. 
In the medium range, performance differences were insignificant. 
However, for large-area changes (80\%-100\%), the unmasked condition was clearly superior, with substantially higher fidelity (2.98 vs. 2.23) and a better \textit{Robustness to New Problems} (2.85 vs. 2.73), with $p\leq0.05$.
While masking could achieve perfect issue resolution on large edits, it often degraded the fidelity by altering unaffected elements.

In summary, the effectiveness of using a mask is highly dependent on the area of the requested change. 
Masking is most effective for small, localized edits where it improves \textit{User Preference Rank} and \textit{Issue Resolution}. 
As the size of the target area increases, the unmasked approach becomes more reliable, as masking offers mixed results for medium-sized changes and is detrimental to UI fidelity for large-area modifications.

\begin{table}[]
\caption{Results of 120 annotations (by 2 annotators for 60 tasks) comparing GPT-Image-1 with and without mask.}

\scalebox{0.85}{
\tabcolsep=0.1cm
\begin{tabular}{cc | c | ccc|c | ccc|c | ccc|c}
\hline
 & & \textbf{F\#1} & \multicolumn{4}{c|}{\textbf{Resolution}} & \multicolumn{4}{c|}{\textbf{Fidelity}} & \multicolumn{4}{c}{\textbf{Robustness}} \\
 & & & \multicolumn{3}{c|}{\textbf{Frequency}} & \textbf{Avg} & \multicolumn{3}{c|}{\textbf{Frequency}} & \textbf{Avg} & \multicolumn{3}{c|}{\textbf{Frequency}} & \textbf{Avg} \\
 & & & \textbf{1} & \textbf{2} & \textbf{3} &  & \textbf{1} & \textbf{2} & \textbf{3} &  & \textbf{1} & \textbf{2} & \textbf{3} &  \\
\hline
\multirow{2}{*}{S} & Mask & \textbf{28} & 6 & 4 & 30 & \textbf{2.60} & 0 & 0 & 40 & \textbf{3.00} & 1 & 10 & 29 & 2.70 \\
                  & No mask & 12 & 8 & 2 & 30 & 2.55 & 0 & 4 & 36 & 2.90 & 0 & 5 & 35 & \textbf{2.88} \\
\hline
\multirow{2}{*}{M} & Mask & \textbf{21} & 0 & 2 & 38 & \textbf{2.95} & 5 & 2 & 33 & 2.70 & 1 & 7 & 32 & 2.77 \\
                  & No mask & 19 & 1 & 2 & 37 & 2.90 & 2 & 6 & 32 & \textbf{2.75} & 0 & 3 & 37 & \textbf{2.92} \\
\hline
\multirow{2}{*}{L} & Mask & 18 & 0 & 0 & 40 & \textbf{3.00} & 11 & 9 & 20 & 2.23 & 1 & 9 & 30 & 2.73 \\
                  & No mask & \textbf{22} & 3 & 1 & 36 & 2.83 & 0 & 1 & 39 & \textbf{2.98} & 0 & 6 & 34 & \textbf{2.85} \\
\hline
\end{tabular}
}

\vspace{0.5em}

\raggedright
\footnotesize
\textbf{S}: Small area of 0\%-20\%;
\textbf{M}: Medium area of 20\%-80\%;
\textbf{L}: Large area of 80\%-100\%;
\textbf{Mask}: Using mask;
\textbf{No mask}: Not using mask;
\textbf{F\#1}: Frequency of achieving rank 1 in \textit{User Preference Rank}; 
\textbf{Frequency} of the scores 1,2,3;
\textbf{Avg}: Average score;
\textbf{Resolution}: \textit{Issue Resolution}; 
\textbf{Fidelity}: \textit{Fidelity to Original}; 
\textbf{Robustness}: \textit{Robustness to New Problems}.

\label{tab:mask-eval}
\end{table}

\subsection{RQ1.3. Ablation Study} \label{sec:prompt-eval}

\likethis has a ``Suggestion Generation'' step designed to generate textual solution specifications based on user feedback. 
However, even without this step, the image generation models can still edit UI images using a simple prompt asking them to address the user feedback.
To evaluate its impact, we conducted an ablation study, removing the ``Suggestion Generation''.

\subsubsection{Evaluation Setting}
For this study, we are using GPT-Image-1 and GPT-Image-1-mask.
The variants lacking ``Suggestion Generation'' are referred as GPT-Image-1-no-sg and GPT-Image-1-mask-no-sg. 
For the ``no-sg'' configurations, we employed Prompt \ref{prompt:ui-edit-no-solution}, which directly incorporates user feedback for UI suggestion generation, instead of the previously used Prompt \ref{prompt:solution} and \ref{prompt:ui-edit}.

\subsubsection{Evaluation Method}
We split the 120 screenshots and corresponding feedback collected in Section \ref{sec:dataset} into two equal sets of 60.
The first set was used for a comparative analysis of the GPT-Image-1 and GPT-Image-1-no-sg, while the second set was used to compare the GPT-Image-1-mask and GPT-Image-1-mask-no-sg. 
Subsequently, two annotators independently annotated these 60+60 tasks with our annotation tool following the evaluation protocol.

\begin{table}[!htp]
\caption{Results of 120 annotations (by 2 annotators for 60 tasks) comparing GPT-Image-1-mask with and without ``Suggestion Generation'', and 120 annotations (by 2 annotators for 60 tasks) comparing GPT-Image-1 with and without ``Suggestion Generation''.}

\scalebox{0.85}{
\tabcolsep=0.1cm
\begin{tabular}{cc | c | ccc|c | ccc|c | ccc|c}
\hline
 & & \textbf{F\#1} & \multicolumn{4}{c|}{\textbf{Resolution}} & \multicolumn{4}{c|}{\textbf{Fidelity}} & \multicolumn{4}{c}{\textbf{Robustness}} \\
 & & & \multicolumn{3}{c|}{\textbf{Frequency}} & \textbf{Avg} & \multicolumn{3}{c|}{\textbf{Frequency}} & \textbf{Avg} & \multicolumn{3}{c|}{\textbf{Frequency}} & \textbf{Avg} \\
 & & & \textbf{1} & \textbf{2} & \textbf{3} &  & \textbf{1} & \textbf{2} & \textbf{3} &  & \textbf{1} & \textbf{2} & \textbf{3} &  \\
\hline
\multirow{2}{*}{Mask} & SG & \textbf{108} & 10 & 10 & 100 & \textbf{2.75} & 5 & 26 & 89 & \textbf{2.70} & 2 & 29 & 89 & \textbf{2.73} \\
                  & N-SG & 12 & 78 & 7 & 35 & 1.64 & 66 & 11 & 43 & 1.81 & 61 & 22 & 37 & 1.80 \\
\hline
\hline
\multirow{2}{*}{No mask} & SG & \textbf{80} & 10 & 11 & 99 & \textbf{2.74} & 1 & 12 & 107 & 2.88 & 0 & 11 & 109 & \textbf{2.91} \\
                  & N-SG & 40 & 25 & 13 & 82 & 2.48 & 3 & 6 & 111 & \textbf{2.90} & 3 & 16 & 101 & 2.82 \\
\hline
\end{tabular}
}

\vspace{0.5em}

\raggedright
\footnotesize
\textbf{SG}: Using ``Suggestion Generation'';
\textbf{N-SG}: Not using ``Suggestion Generation'';
\textbf{Mask}: Using mask;
\textbf{No mask}: Not using mask;
\textbf{F\#1}: Frequency of achieving rank 1 in \textit{User Preference Rank}; 
\textbf{Frequency} of the scores 1,2,3;
\textbf{Avg}: Average score;
\textbf{Resolution}: \textit{Issue Resolution}; 
\textbf{Fidelity}: \textit{Fidelity to Original}; 
\textbf{Robustness}: \textit{Robustness to New Problems}.

\label{tab:prompt-eval}
\end{table}

\subsubsection{Evaluation Result}

Table \ref{tab:prompt-eval} demonstrates a clear advantage for incorporating Suggestion Generation (SG).
For the 60 tasks with masks, agreement of two annotators was 91.6\%, while for the 60 tasks without masks, it was 89.58\%.

With masks, the SG setting consistently outperformed the no-SG setting. 
The SG setting had significantly better \textit{User Preference Rank} (108 vs. 12), it also showed superior \textit{Issue Resolution} (2.75 vs. 1.64), better \textit{Fidelity to Original} (2.70 vs. 1.80), and stronger \textit{Robustness to New Problems} (2.73 vs. 1.80), all results with $p\leq0.05$.

Without masks, SG's superior performance largely continued. 
The SG setting again achieved a better \textit{User Preference Rank} (80 vs. 40) and a higher scores in \textit{Issue Resolution} (2.74 vs. 2.48), both results with $p\leq0.05$.
The only exception was \textit{Fidelity to Original}, where the no-SG setting scored slightly higher at 2.90, suggesting it made marginally fewer changes to untargeted UI elements.

\noindent
\ul{In summary, these results clearly confirm the importance of the ``Suggestion Generation'' step in our approach}.

\section{Users' and Developers' Study}
\label{sec:user-study}
To answer RQ2, we conducted user testing with 10 real apps and then surveyed the corresponding developers.
The user testing evaluated whether improvement suggestions generated with \likethis accurately captured participants’ issues and whether users considered the approach helpful to submit feedback (RQ2.1).
The developer survey examined whether adding improvement suggestions increased the understandability and actionability of the feedback compared to reports without suggestions (RQ2.2).

\subsection{Study Design}

\begin{table}
    \centering
    \caption{Participants in user and developer studies.}

    \setlength{\tabcolsep}{4pt}
    \renewcommand{\arraystretch}{1.05}

    \begin{tabularx}{\linewidth}{|c|c|c|X|}
        \hline
        \multicolumn{4}{|c|}{\textbf{Users}} \\
        \hline
        Id & Gender & Age & Occupation \\ [0.5ex]
        \hline
        U1 & M & 28 & Dentist \\
        U2 & M & 19 & Pediatric caregiver \\
        U3 & M & 27 & Marketing consultant \\
        U4 & F & 26 & Teacher \\
        U5 & F & 28 & Kindergarten teacher \\
        U6 & F & 30 & Fitness trainer \\
        U7 & M & 29 & Industrial technician \\
        U8 & F & 28 & Doctor \\
        U9 & M & 28 & Civil Engineer \\
        U10 & M & 32 & Software Engineer \\
        U11 & M & 27 & Lifeguard \\
        U12 & M & 27 & Media Studies student \\
        U13 & M & 32 & Retail salesman \\
        U14 & F & 30 & Automotive Sales Consultant \\
        U15 & M & 33 & Construction Manager \\
        \hline
    \end{tabularx}
    \vspace{0.4em}

    \begin{tabularx}{\linewidth}{|c|c|c|c|>{\raggedright\arraybackslash}X|>{\raggedright\arraybackslash}X|}
        \hline
        \multicolumn{6}{|c|}{\textbf{Developers}} \\
        \hline
        Id & G & Age & E & Role & App Categories \\ [0.5ex]
        \hline
        D1 & M & 55 & 25 & Freelance Mobile Developer & Weather, Entertainment \\
        \hline
        D2 & M & 41 & 13 & Senior Lead IT Architect & Games, Finance \\
        \hline
        D3 & M & 30 & 7 & Independent Mobile Developer & Graphics \& Design, Health \& Fitness \\
        \hline
        D4 & M & 29 & 5 & Application Architect & Reference, Utilities \\
        \hline
        D5 & F & 30 & 5 & iOS Developer & Productivity, Photo \& Video \\
        \hline
    \end{tabularx}

    \begin{minipage}{\linewidth}
        \vspace{0.1cm}
        \footnotesize
        \textbf{G}: Gender, \textbf{E}: Years of Experience.
    \end{minipage}
    \label{table:userTestingParticipants}
\end{table}

We conducted user testing with 15 participants and 10 apps.
The testing results were later shared with the developers of the 10 apps to gather their feedback on what users reported.

\subsubsection{Participant Recruitment}
To qualify for our study, we looked for developers in our personal network with at least five years of experience and who have maintained two apps for at least half of their history, ensuring significant experience and strong familiarity with the apps.
We identified 16 developers who were accessible to us and who met those criteria.
We contacted five who all agreed to participate: four men and one woman.
The 10 corresponding apps cover 10 different domains and categories ensuring a diverse setting.
For user recruitment, we reached out to 20 volunteers through private and professional online groups, of whom 15 agreed to participate. 
To diversify perspectives, we specifically sought participants with non-technical backgrounds (e.g., dentist, lifeguard).
Details of all participants are summarised in Table~\ref{table:userTestingParticipants}.
All participants provided informed consent and there was no compensation for participation. 
The study followed our institution’s Ethics Committee guidelines.

\subsubsection{User Testing Procedure}
We conducted user testings with 15 participants in convenient locations such as homes or offices.
Each participant was requested to interact with two randomly selected apps from the 10 apps.
They were given a list of tasks designed to guide them through the main features of each app, following a common task-based usability testing approach~\cite{nielsen1994usability}.
In the end, each app was tested by three different participants.
Each session lasted for $\sim$45 minutes.

During user testing, participants were asked to identify and report with \likethis any UI issues they encountered in the test apps.
To streamline the process, each participant was given an iPhone 16 preloaded with \likethis and the 10 apps.
The device was customized using Apple Shortcuts \cite{appleShortcut}, App Intents \cite{appintents} and Accessibility Settings \cite{appleAccessibility}: a triple tap on the back of the phone took a screenshot and automatically opened \likethis with the screenshot preloaded.
This setup ensured that the feedback process was easily accessible to participants, serving as a workaround since we could not integrate our framework directly into the production apps.
In practice, \likethis functionality could be triggered through a simple ``report issue'' button inside an app.
We recorded the screen and voice to capture all interactions for later analysis.
Throughout the sessions, one author sat beside the participants with a laptop, ready to answer clarification questions and take observational notes.

The sessions followed four steps.
First, we welcomed the participants, explained the purpose of the study, informed them of their right to opt out at any time, and obtained their participation consent.
Second, we demonstrated the feedback submission process, and introduced the think-aloud protocol to be used \cite{nielsen1994usability}.
We stressed that their task was to use the apps naturally but report any UI issue they encounter.  
To avoid bias towards AI, we did not disclose the underlying AI pipeline.
Third, participants performed tasks on the provided iPhone, reporting any issues they encountered while using think-aloud, as we observed and took notes.
After each issue submission, participants answered a follow-up question about how accurately their selected improvement suggestion addressed their issue.
Finally, we conducted a short exit survey asking how helpful a feedback mechanism with improvement suggestions was and captured overall impressions and comments.

\subsubsection{Developer Survey Procedure}

After collecting improvement suggestions from the user testings, we aimed to understand whether these are more helpful for developers to understand and act on the feedback.
For this, we distributed questionnaires to five developers, each from a different development team responsible for one of the 10 apps.
Each question focused on a single user issue and comprised three pages:
On the first page, the issue was presented alongside a screenshot of the original screen and, if available, the marked area.
Developers then assessed how understandable and actionable the reported issue was on its own.
On the second page, the same issue was shown again, this time including the improvement suggestion and if available, the user comment.
Developers then assessed whether the improvement suggestion increased the understandability of the issue and the actionability of the feedback.
On the last page, developers answered open-ended questions, reflecting on the \likethis approach and sharing any concerns or suggestions.

\subsection{RQ2.1. User Testing Results}

\begin{table}
\caption{``How accurate is your selected suggestion in addressing the issue you wanted to report?'' (User testing)}
\centering
    \scalebox{0.9}{
    \begin{tabularx}{1.1\linewidth}{X|c|c}
        \hline
        \textbf{Rating} & \textbf{Count} & \textbf{Percentage} \\
        \hline
        Very inaccurate & 0 & 0\% \\
        Rather inaccurate & 0 & 0\% \\
        Somewhat accurate & 5 & 6.9\% \\
        Mostly accurate & 22 & 30.6\% \\
        Very accurate & 41 & 56.9\% \\
        \hline
        Canceled submission & 4 & 5.6\% \\
        \hline
        \textbf{Total} & 72 & 100\% \\
        \hline
    \end{tabularx}
    }
\label{tab:user-survey-results-q1}
\end{table}

In total, users submitted 68 improvement suggestions for the 10 apps, while four suggestions were canceled before submission, indicating that none of the generated improvements were suitable.
Table~\ref{tab:user-survey-results-q1} shows how users rated the accuracy of their chosen suggestions in addressing their issues.
Overall, the responses were very encouraging: 56.9\% of the 68 submitted suggestions were rated as \textit{Very accurate}, and 30.6\% as \textit{Mostly accurate}.
Only 6.9\% were rated as \textit{Somewhat accurate}, and none were rated as \textit{Rather inaccurate} or \textit{Very inaccurate}.

During think-aloud, participants often said that the suggestions matched or exceeded their expectations, sometimes describing them as ``exactly what I was thinking but better.''
However, participants sometimes rated a suggestion as \textit{Mostly accurate} rather than \textit{Very accurate} when they were unsure whether it was the best possible solution or when the suggestion introduced new issues.
For instance, some noted that the suggestions looked ``ugly'', or that the proposed change would fix the issue but cause side effects elsewhere in the UI, such as removing buttons.
Among the four cases where users canceled the submission by selecting ``I like none'', two found that none of the generated suggestions solved their issues, while the other two felt misunderstood by the system, as the suggestions focused on something else.

\begin{table}
\caption{``When encountering issues with the apps you regularly use, do you think this feedback approach (with suggestions to select from) will help you to submit your complaints and feedback?'' (User testing)}
\centering
    \scalebox{0.9}{
    \begin{tabularx}{1.1\linewidth}{X|c|c}
        \hline
        \textbf{Response} & \textbf{Count} & \textbf{Percentage} \\
        \hline
        Strongly disagree & 0 & 0\% \\
        Disagree & 1 & 7.1\% \\
        Undecided & 1 & 7.1\% \\
        Agree & 5 & 35.7\% \\
        Strongly agree & 8 & 57.1\% \\
        \hline
        \textbf{Total} & 15 & 100\% \\
        \hline
    \end{tabularx}
    }
\label{tab:user-survey-results-q3}
\end{table}

Table~\ref{tab:user-survey-results-q3} shows how the 15 participants rated whether they think a feedback approach offering improvement suggestions would help them submit complaints and feedback in the apps they regularly use.
The responses were largely positive: eight people selected \textit{Strongly agree}, and five picked \textit{Agree}.
Only one person was \textit{Undecided}, and another selected \textit{Disagree}, while no participants \textit{Strongly disagreed}.

When participants explained their experiences, a central theme was that they felt their communication improved when using\newline\likethis.
Many reported struggling to explain UI issues precisely, often due to a lack of technical vocabulary.
One participant said, ``I know what’s bothering me, but I don’t know how to say it. Improvement suggestions feel like a translation of my problem into something a developer might understand well.''
Others described the suggestions as confirmation that they had expressed the issue clearly:
``When the generated suggestion made sense, I knew I had explained the problem clearly.''
This quote also reflects the fact that most participants immediately recognized the suggestions as AI-generated.

Despite the overall positive impressions, concerns were also raised.  
Some participants felt that the additional interactions could be burdensome, as one said, ``I want to hand in my problem as fast as possible.''
The two people who selected \textit{Undecided} and \textit{Disagree} both did so because, as one explained, ``I don't know if this is helpful for me; I feel like it is mostly for the developers.''

Taken together, these reflections directly address RQ2.1. Users perceived the GenAI-generated suggestions as highly accurate in addressing their reported issues. 
They found the concept of \likethis helpful for making their feedback clearer, though some expressed concerns about the added effort.

\subsection{RQ2.2. Developer Survey Results}

Table~\ref{tab:developer-survey-understandability} shows how developers rated the understandability of user feedback before and after seeing associated improvement suggestions.
For the 68 reports, when developers only saw the original user issue along with a screenshot (and optionally marked area), 70.6\% were rated as \textit{clearly understandable}, 22.1\% as \textit{somewhat understandable}, and 7.4\% as \textit{not understandable}.
After seeing the associated improvement suggestion, 27.9\% of issues were rated as \textit{more understandable than before}, 70.6\% as \textit{about the same}, and only 1.5\% as \textit{less understandable}.
A one-sided paired sign test showed a statistically significant difference ($p\leq0.05$), indicating that improvement suggestions improve developers' understanding of user feedback.

\begin{table}
\caption{Understandability of user feedback before and after seeing improvement suggestions. (Developer survey)}
\centering
    \scalebox{0.9}{
    \begin{tabularx}{1.1\linewidth}{X|c}
        \hline
        \textbf{Before seeing the improvement suggestion} & \textbf{Count (\%)} \\
        \hline
        Not Understandable & 5 (7.4\%) \\
        Somewhat Understandable & 15 (22.1\%) \\
        Clearly Understandable & 48 (70.6\%) \\
        \hline
        \textbf{Total} & 68 (100\%) \\
        \hline
    \end{tabularx}
    }

\vspace{0.5em}
    \scalebox{0.9}{
    \begin{tabularx}{1.1\linewidth}{X|c}
        \hline
        \textbf{After seeing the improvement suggestion} & \textbf{Count (\%)} \\
        \hline
        Less Understandable than before & 1 (1.5\%) \\
        About the Same as before & 48 (70.6\%) \\
        More Understandable than before & 19 (27.9\%) \\
        \hline
        \textbf{Total} & 68 (100\%) \\
        \hline
    \end{tabularx}
    }
\label{tab:developer-survey-understandability}
\end{table}

\begin{table}
\caption{Actionability of the textual feedback (A) and the associated improvement suggestions (B). (Developer survey)}
\centering
    \scalebox{0.9}{
    \begin{tabularx}{1.1\linewidth}{X|c|c}
        \hline
        \textbf{Response} & \textbf{A} & \textbf{B} \\
        \hline
        Yes, I could directly implement a change the suggestion & \centeringcell{40\\(58.8\%)} & \centeringcell{48\\(70.6\%)} \\
        \hline
        Possibly, but I would need more information & \centeringcell{15\\(22.1\%)} & \centeringcell{17\\(25.0\%)} \\
        \hline
        No, I don't know what I should do & \centeringcell{13\\(19.1\%)} & \centeringcell{3\\(4.4\%)} \\
        \hline
        \textbf{Total} & \multicolumn{2}{c}{68 (100\%)} \\
        \hline
    \end{tabularx}
    }
\label{tab:developer-survey-actionability}
\end{table}

Table~\ref{tab:developer-survey-actionability} shows how developers rated the actionability of reported feedback compared to the actionability of associated improvement suggestions.
When evaluating issues alone, developers selected \textit{yes, I could directly implement a change} in 58.8\% of cases and \textit{no, I don't know what to do} in 19.1\%.  
Evaluating the improvement suggestions, \textit{yes} rose to 70.6\% (+11.8\%) and \textit{no} dropped to 4.4\% ($-14.7$\%), indicating that improvement suggestions tend to be rated more actionable.
A one-sided paired sign test showed a statistically significant increase in actionability of improvement suggestions compared to issues alone ($p\leq0.05$).

When explaining the ratings, developers repeatedly described the feedback with improvement suggestions as engaging and enlightening.
One developer said: ``Going through the suggestions was genuinely fun, and I was surprised by solutions I wouldn’t have thought of.''
3 of 5 noted that the improvement suggestions clarified the issues that were initially rather vague or hard to interpret.
One said: ``The suggestion gave me a clearer idea of what the user actually wanted, especially when the original report was unclear.''

However, one cautioned: ``Before I see a fix, I want to dive deeper into the root cause first.''
Another one said: ``These proposals focus on a single screen and overlook how a change might affect the rest of the user journey.
In some cases, the real fix belongs two steps earlier.''
Feasibility was another concern. 
All highlighted that some suggestions break platform rules or ignore important edge cases, so a validation step is still required before implementation.

These findings address RQ2.2, indicating that feedback augmented with generated improvement suggestions leads to a statistically significant improvement in developers' understanding and the actionability of issues.
However, developers also stressed the importance of reviewing suggestions critically, as these may overlook underlying causes, broader app context, or feasibility constraints.

\section{Limitations and Threats to Validity}
\label{sec:threats}
As with every empirical work, ours also has several potential threats to validity, which we briefly discuss in the following. 

\textit{Subjectivity in Manual Annotation.}
During the benchmarking of four state-of-the-art GenAI models, we performed extensive manual assessments and comparisons of generated UIs. 
Acknowledging that subjectivity and potential bias are inherent in any manual labelling, we implemented several mitigation strategies.
First, we developed a custom annotation tool that randomised and anonymised the order in which different UI suggestions were displayed, making it practically impossible for the evaluator to guess, which image is generated by which model or which configuration. 
Second, to minimize human errors, every UI suggestion was evaluated independently by two annotators. 
Each annotator has more than five years of professional development experience, which was important for evaluating the criteria of \textit{Issue Resolution}, \textit{Fidelity to Original}, and \textit{Robustness to New Problems}.
Third, to ensure consistency, we developed an annotation guide, including a well-defined semantic scale for each scoring criterion. 
We also conducted a pre-annotation pilot for 15 tasks followed by a meeting to discuss and calibrate the interpretations.
Those measures led to high agreement rates $>80\%$. 
Still, in some cases, evaluators disagreed on rank (agreement rate = 75.5\%), resolution (87.6\%), fidelity (87.7\%), and robustness (83.6\%).
In 77\% of those disagreements, the annotators' scores differed by one point only.
This reflects an inherently subjective component in UI assessment as different users might also disagree. 
We thus report on the frequencies by both annotators in addition to the averages.
While annotators were experienced developers, they did not know the annotated screens.
Thus, we think that a potential developer bias for \textit{User Preferences Rank} is rather marginal.

\textit{Designer vs. User Issue Description.}
The issues from the UICrit dataset were authored by designers, which may not fully reflect how end-users articulate issues.
Our analysis on a subset of the UICrit dataset and UISMiner dataset \cite{Wang:UISMinerMiningUI:2022} (a dataset of UI-related user reviews) reveals that, for the similar underlying issue, designer-authored issue descriptions (e.g., ``the buttons are too close together, which may lead to difficulty in distinguishing between them and accidental clicks'') generally align more with design guidelines~\cite{Nielsen:EnhancingExplanatoryPower:1994,Luther:StructuringAggregatingEvaluating:2015,apple-design-guideline} than user-authored ones (e.g., ``Please make the magnet button not too close to the play button, I get frustrated when I accidentally click [it]''). 
Nevertheless, our pilot testing revealed that, once processed through the ``Suggestion Generation'' component, issues with different phrasing styles tend to yield very similar ``Improvement Suggestions''. 
Currently unavailable UI datasets annotated with feedback by actual users would certainly enable replicating our benchmarking study and strengthening its validity.  
Finally, the benchmarking experiments were designed for comparative analyses of different models and settings. 
Since the models received identical inputs, the comparative evaluation results remain valid despite the variation in issue-description styles.

\textit{Influence of Controlled Environment.}
In our study, users were asked to report issues they encountered while completing predefined tasks in apps they might be unfamiliar with.
Also, they were encouraged to think aloud to make their thoughts easier to observe.
Both factors may have influenced their feedback behaviour.
Despite us not mentioning AI, most participants noticed that suggestions were generated with AI.
This might have introduced a bias, on when to submit an issue and how to write the feedback.
For instance, users might try to phrase feedback based on their experience with AI bots.
Participants used a prepared iPhone with \likethis configured, where a triple-tap on the back triggered the feedback process. In a real environment, users might behave differently based on how our approach is implemented, for example via a ``report issue'' button.
All participants saw the generation approach of \likethis at the beginning, as we decided against a separate control group with a conventional text-only feedback interface. 
Such experimental design would have reduced/halved the number of generated improvements in our dataset. 
Moreover, as each feedback is rather unique, it is hard to compare different reported issues across different groups. 
We let users identify issues on their own, rather than providing predefined ones, as predefined issues would have restricted natural feedback behaviour and made the study less reflective of real-world scenarios, reducing realism.
When developers first assessed the ``issue only'' condition, they saw these issues exactly as users wrote them. 
This wording might already been shaped by the users behaviour and expectation towards AI.
These factors might limit how well our baseline (original feedback) reflects real user feedback compared to what is seen in app stores. 
Long-term field studies for apps with large user bases will likely lead to diverse feedback, more robust results, and additional insights, but requires strong commitment by vendors and performance tuning to reduce image generation time.

\textit{Sample Size and Participants Recruitment.}
For the user and developer study, our main focus was \textit{realism}. 
That is, we aimed to capture thoughts of different users reporting real feedback, while using real production apps, and how the actual developers assess this feedback.
By studying 10 production apps from different domains and of different complexity levels, we ensured variation of contexts.
At the same time, the limited number of participants (15 users and 5 developers) might restrict the generalisability of the findings. 
In fact, we refrain from claiming generalisability of the reported sentiments or behaviours as examining behavioural patterns would require a much larger sample.
However, we consider the sample size of 68 feedback reports appropriate for our comparative analysis on the impact of \likethis, providing sufficient statistical power to detect meaningful differences. 
This aligns with similar prior work, such as by Cutler et al.~\cite{cutler2025crowdsourced} and Pourasad and Maalej~\cite{Pourasad:DoesGenAIMake:2025}, who each recruited 10 participants for their user testing studies.

As we recruited participants from our network, selection, participation, and social-desirability biases remain possible. 
Similar to user testing methods such as usability testing \cite{natesan2016cognitive}, these biases may have influenced the willingness to participate and engage during testing, potentially leading to higher quantity and quality of reports, as well as more positive survey ratings.
Additionally, while users covered a variety of occupations mostly unrelated to IT, the age range of participants was narrow (adults up to 33) and only one-third were female.
Nevertheless, we still believe the mixed sentiments observed from users and developers make our core findings robust. 
Future replications with more diverse users, apps, and developers could further enhance generalisability.

\section{Related Work}
\label{sec:rel-work}

\subsection{User Feedback in Software Development}
User feedback is a crucial source of information for requirements engineering, UI design, and software evolution in general \cite{Maalej:AutomatedProcessingUser:2025}.
Feedback is nowadays largely available and easily accessible in diverse channels such as app stores \cite{Pagano:UserFeedbackAppstore:2013}, X (formerly Twitter) \cite{Nayebi:AppStoreMining:2018}, or Reddit \cite{Iqbal:MiningRedditNew:2021}.
Over the last decade, researchers proposed AI-based approaches to automatically analyse, classify, and summarise feedback \cite{dkabrowski:EMSE:2022} for informing development decisions. Only recently researchers started to highlight the importance of feedback quality \cite{Maalej:AutomatedProcessingUser:2025}, e.g. by augmenting it with data collected in the background. 

\subsubsection{User Feedback Analysis}

Given the large volume of user feedback, manual analysis is impractical. 
To address this challenge, a variety of automated techniques have been developed to efficiently extract valuable information, thereby reducing the manual effort required.
These techniques include classification, clustering, summarization, feature extraction, etc.:
Classification involves categorizing user feedback into predefined groups such as feature requests or bug reports \cite{Maalej:AutomaticClassificationApp:2016,Mekala:ClassifyingUserRequirements:2021,Wei:ZeroshotBilingualApp:2023,Prenner:MakingMostSmall:2021}.
Clustering groups together user feedback that discuss similar topics by computing semantic similarity \cite{Stanik:UnsupervisedTopicDiscovery:2021,Wang:WhereYourApp:2022,Wei:ZeroshotBilingualApp:2023,Devine:WhatsClusterSoftware:2022}.
Summarization creates a succinct summary for a group of user feedback \cite{Alshangiti:HierarchicalBayesianMultikernel:2022,Gao:ListeningUsersVoice:2022,Wei:ZeroshotBilingualApp:2023}.
Feature extraction focuses on identifying and extracting app features mentioned within user feedback \cite{Johann:SAFESimpleApproach:2017,Wu:IdentifyingKeyFeatures:2021,Assi:LLMCureLLMbasedCompetitor:2025,Motger:LeveragingEncoderonlyLarge:2025}.
While such automated methods can help developers process feedback at scale, e.g. to understand emerging and common topics, our work focuses on the single feedback items.
\likethis empowers users to express their thoughts more precisely by generating several improvements alternatives from which they can choose which one fits best to their perceptions.

\subsubsection{User Feedback Augmentation}

A large portion of textual feedback is shared by non-technical users, who do not necessarily 
understand the information required by developers.
To help developers precisely understand and react to the user needs, feedback can be improved using various augmentation techniques.
Oriol et al.~ \cite{Oriol:FAMESupportingContinuous:2018} proposed a framework for the combined and simultaneous collection of feedback and monitoring data to support continuous requirements elicitation.
Martens and Maalej \cite{Martens:ExtractingAnalyzingContext:2019} proposed a method for extracting basic context information (such as platform, device, application version, and system version) from unstructured, informal feedback. 
If this information needed by developers is missing, a bot can ask the user for clarification or consent to retrieve and submit it.  
Stanik et al.~\cite{Stanik:WhichAppFeatures:2020} utilized the user interaction events for training a machine learning model to learn app feature usage ultimately matching what users say with what they do.
Li et al.~\cite{Li:TellingUsYour:2022} investigated eye movement patterns to evaluate user satisfaction with respect to six non-functional requirements, thereby exposing latent user needs.
Integrating these augmentation strategies with textual feedback can substantially improve developers’ understanding of the context in which the feedback emerged. 
Our work is complementary, as it uses GenAI to focus on \textit{solution} ideas from users that address the encountered issues rather than their context.

\subsection{UI Generation}
Studies which explored UI generation with GenAI, can be divided into two categories: UI image generation and UI code generation.

\subsubsection{UI Image Generation}
Image generation models, such as Diffusion Models \cite{Rombach:HighResolutionImageSynthesis:2022} and Generative Adversarial Networks (GANs) \cite{Goodfellow:GenerativeAdversarialNetworks:2020}, have been widely adopted for design exploration in the context of mobile UIs. 
Mozaffari et al.~proposed GANSpiration \cite{Mozaffari:GANSpirationBalancingTargeted:2022}, a model that takes a UI image as input and generates design suggestions for both targeted and serendipitous inspiration.
More recently, building upon LayoutDM \cite{Inoue:LayoutDMDiscreteDiffusion:2023} and Stable Diffusion \cite{Rombach:HighResolutionImageSynthesis:2022}, Wei et al.~proposed UI-diffuser \cite{Wei:BoostingGUIPrototyping:2023}, a model fine-tuned on Rico \cite{Deka:RicoMobileApp:2017} dataset, for generating UI images conditioned on given textual descriptions and specified UI elements. 
Subsequently, Wei et al.~introduced UI-diffuser-v2 \cite{Wei:AIInspiredUserInterface:2025}, a model trained on 135k screenshot-caption pairs from their SCapRepo dataset \cite{Wei:GUingMobileGUI:2025}, enabling the generation of UI images from textual descriptions alone.
Despite their usefulness, these image-based methods only produce new UIs instead of editing existing ones, making them more suitable for brainstorming and design tasks rather than for creating UI suggestions by users.

\subsubsection{UI Code Generation}
With the increasing capabilities of LLMs, the automatic generation of UI code has become a popular area of research. 
Wu et al.~\cite{Wu:UICoderFinetuningLarge:2024} and Feng et al.~\cite{Feng:DesigningLanguageWireframing:2023} fine-tuned LLMs to generate \texttt{SwiftUI} implementations or HTML-based wireframe from textual descriptions.
Wei et al.~\cite{Wei:AIInspiredUserInterface:2025} and Kolthoff et al.~\cite{Kolthoff:GUIDELLMDrivenGUI:2025} explored using LLMs to decomposes high-level UI descriptions into fine-grained UI section descriptions, which are subsequently translated into Material Design UIs \cite{material_design} or Figma prototypes.
Beyond generation, LLMs have also been leveraged for UI code editing. 
For example, Yuan et al.~proposed DesignRepair \cite{Yuan:DesignRepairDualStreamDesign:2025}, a dual-stream, design-guideline-aware system that analyzes and repairs Material Design UIs at the code level and on the rendered page.
Additionally, Lu et al.~presented Misty \cite{Lu:MistyUIPrototyping:2025}, a workflow enabling developers to efficiently incorporate varying design elements from existing examples into ongoing React-based UI page \cite{react}.
Recently, several commercial tools, such as Stitch \cite{stitch}, Uizard Autodesigner \cite{uizard}, and Figma Make \cite{figma_make}, have been released for UI code generation.
Previous works have investigated both the generation and editing of UI source code using LLMs.
However, existing approaches are framework-dependent (e.g., React or Material Design), which constrains their applicability in modern apps that employ diverse and sophisticated technology stacks.
Moreover, while these approaches aim to assist developers and designers with technical background, our work aims at empowering average users to submit precise suggestions to improve the app and its UI.

\section{Summary and Road Ahead}
\label{sec:conlusion}

Because users often lack expertise, time and skills to precisely articulate their needs and thoughts \cite{Zowghi:EASE:2015}, and due to the inherent ambiguity of textual feedback \cite{Maalej:WhenUsersBecomeCollaborators:2009}, 
research has repeatedly highlighted that user feedback in practice is often vague or uninformative to developers \cite{Pagano:UserFeedbackAppstore:2013, Martens:ExtractingAnalyzingContext:2019}.
Yet, some users do invest effort to provide detailed feedback \cite{Pagano:UserFeedbackAppstore:2013}.
We propose \likethis, a novel GenAI-based approach to help app users submit higher-quality feedback, particularly with respect to accuracy, understandability, and actionability for developers.
\likethis generates multiple improvement suggestions to an existing design that potentially address UI-related issues the moment they are encountered by users, allowing them to select their preferred ``fix''.
A benchmarking study of four GenAI models with a users' and developers' study on 10 real apps indicates that \likethis can effectively empower users to submit accurate improvement suggestions that are understandable and actionable to developers. 
We observed that GPT-Image-1 outperforms other models, that marking small areas ``to fix'' improves the fidelity to the original UI, and most importantly, that an intermediate step to generate specifications of the improved designs before prompting the model to modify the original UI is key to the overall performance. 
While developers in our study appreciated and were even surprised by the novelty of some suggestions, they also highlighted important considerations before implementing the suggested changes, such as platform constraints.
Moreover, whether and when certain users would value the ``expected return on investment of their additional feedback effort'' remains an open question for future studies. 

We argue that GenAI approaches, like \likethis, will fundamentally change not only how users share feedback, but also how they interact and possibly collaborate with developers \cite{Maalej:WhenUsersBecomeCollaborators:2009}. 
GenAI can \textit{translate} ambiguous feedback into improvement suggestions or even source code and patches, to then ``translate'' back technical decisions to users. 
We thus envision several future directions:

\textbf{Swifter trusted generations}: 
Users expressed concerns with the waiting time for the UI generation. 
Current models as GPT-Image-1 need $\sim$1 minute per generation, likely too long for a quick feedback.  
Research into faster models, trained specifically on app usage and UI data, would increase the applicability and acceptance of such approaches.
Those models should carefully protect the user data (e.g.~ blurring their inputs) and ideally run on their devices. 

\textbf{From single to multiple screens:} 
Our current approach focuses solely on single-screen issues. 
However, issues might arise across multiple screens. 
Future research should address this by supporting issue reporting and idea generation that span multiple screens covering the entire user journey and experience.

\textbf{Generating user patches:} 
Our current work empowers users to generate visual mockups. 
However, there could be an immense value in extending it to generate corresponding code changes or even patches that can instantly be deployed. 
While this would boost personalization and potentially speed-up software evolution and maintenance, it also brings major research challenges to ensuring software quality, consistency, and security.

\textbf{Multimodal feedback with explanation:} 
Visual feedback with \likethis and potentially other forthcoming methods will transform how user feedback should be shared and analyzed. 
The users' \textit{expectations} will also likely change, as they would expect to know ``what happened to their feedback''. 
Such explanation (feedback to users) could even be generated at the time of feedback submission, e.g., if the issue is known or if it's rather their mistake.

\begin{acks}
The first author was funded by the Claussen-Simon-Stiftung (PostDoc Plus programme).
We thank all study participants for their time and feedback.    
\end{acks}

\bibliographystyle{ACM-Reference-Format}
\bibliography{ref}

\end{document}